\documentclass[12pt]{article}
\usepackage{graphicx, mathrsfs,amsmath,amsfonts,,bm,float}

\begin{document}

\title{\vskip.5cm
Approximate closed-form formulas for the zeros of the Bessel Polynomials}
\author{Rafael G. Campos and Marisol L. Calder\'on\\
Facultad de Ciencias F\'{\i}sico-Matem\'aticas,\\
Universidad Michoacana, \\
58060, Morelia, Mich., M\'exico.\\
\hbox{\small rcampos@umich.mx, mlopez@fismat.umich.mx}\\
}
\date{}
\maketitle
{\vskip1cm
\noindent MSC: 33C47, 33F05, 33C10, 30C15
}\\
\vspace*{1truecm}

\noindent{\large\bf Abstract}
\vspace*{.5cm}
\\
We find approximate expressions $\tilde{x}(k,n)$ and $\tilde{y}(k,n)$ for the real and imaginary parts of the $k$th zero $z_k=x_k+iy_k$ of the Bessel polynomial $y_n(x)$. To obtain these closed-form formulas we use the fact that the points of well-defined curves in the complex plane are limit points of the zeros of the normalized Bessel polynomials. Thus, these zeros are first computed numerically through an implementation of the electrostatic interpretation formulas and then, a fit to the real and imaginary parts as functions of $k$ and $n$ is obtained. It is shown that the resulting complex number $\tilde{x}(k,n)+i\tilde{y}(k,n)$ is $O(1/n^2)$-convergent to $z_k$ for fixed $k$.
\vskip1.5cm
\section{Introduction}
The polynomial solutions of the differential equation
\begin{equation}\label{edobspo}
x^2y''(x)+2(x+1)y'(x)-n(n+1)y(x)=0
\end{equation}
were studied systematically in \cite{Kra49} by the first time. They are named Bessel polynomials and are given explicitly by
\begin{equation}\label{bespoy}
y_n(x)=\sum_{k=0}^n\frac{(n+k)!}{(n-k)!k!}\left(\frac{x}{2}\right)^k,
\end{equation}
where $n=0,1,\cdots$. Many properties as well as applications are associated to these polynomials: the solution of the wave equation in spherical coordinates, network and filter design, isotropic turbulence fields, and more (see the monograph \cite{Gro78} or \cite{Sri84}-\cite{Bia07} and references therein for some other results). Among these, several results about the important problem concerning the location of its zeros have been obtained \cite{Car92}-\cite{Bru81}. Just for instance, the use of the reverse Bessel polynomials $\theta_n(x)=x^n y_n(1/x)$ in filter design is known since the time when these polynomials began to be studied \cite{Gro78, Bia07}. Here, the poles of the transfer function are essentially the zeros of $\theta_n(x)$. Thus, it is desirable to acquire new analytical knowledge about the location of the zeros of the Bessel polynomials.\\ 
In this note we give approximate explicit formulas for both the real and imaginary parts of the $k$th zero $z_k=x_k+iy_k$ of $y_n(x)$ and show that the approximation order of these new formulas to the exact zeros of the Bessel polynomials is $O(1/n^2)$ for fixed $k$.\\
The approach followed in this note is simple and based on three items. The first is the electrostatic interpretation of the zeros of polynomials satisfying second order differential equations \cite{Sze75}-\cite{Cam99}, the second is a simple curve fitting of numerical data and the third is the known fact that the points of well-defined curves in the complex plane are limit points of the zeros of the normalized Bessel polynomials \cite{Car92}-\cite{Bru81}. The formulas yielded by the electrostatic interpretation of the zeros of Bessel polynomials are used to find them numerically as it has been done previously with these and other sets of points \cite{Pas00}-\cite{Cam95}. Several sets of zeros are computed in this way and the sets of real and imaginary values are fitted by polynomials depending on the index $k$ whose coefficients depend on $n$. Finally, it is found that the approximate expression for the $k$th zero of $y_n(x)$ is $O(1/n^2)$-convergent to a limit point of the zeros of the Bessel polynomials. Since the exact zero $z_k$ is $O(1/n^2)$-convergent to its limit point \cite{Car92}, we conclude that the approximation order of our approximate expression to the exact $k$th zero is also $O(1/n^2)$.
\section{Asymptotic expressions for the zeros}\label{asymex}
Let $z_k=x_k+iy_k$, $k=1,2,\cdots,n$, be the zeros of the Bessel polynomial $y_n(x)$. Then, from (\ref{edobspo}) follows that
\[
\sum_{k=1}^N\frac{1}{z_j-z_k}+\frac{(z_j+1)}{z_j^2}=0, 
\]
where $j=1,2,\cdots,n$, i.e., the real and imaginary parts of the zeros should satisfy the electrostatic equations
\begin{eqnarray}\label{elein}
\sum_{k=1}^N\frac{x_j-x_k}{(x_j-x_k)^2+(y_j-y_k)^2}+\frac{x_j^3+x_j^2+x_j y_j^2-y_j^2}{\left(x_j^2+y_j^2\right)^2}&=&0,\\
\sum_{k=1}^N\frac{y_j-y_k}{(x_j-x_k)^2+(y_j-y_k)^2}+\frac{y_j \left(x_j^2+2 x_j+y_j^2\right)}{\left(x_j^2+y_j^2\right)^2}&=&0.\nonumber
\end{eqnarray}
This set of nonlinear equations can be solved by standard methods. We have used a Newton method to solve them up to $n=500$. As it is shown in Fig. 1, the piecewise linear interpolation of the real and imaginary parts of the zeros of the normalized Bessel polynomials $y_n(x/n)$ can be fitted by polynomials of the second and third degree in the index $k$. 
\begin{figure}[H]\label{Figuno}
\centering
\includegraphics[scale=0.55]{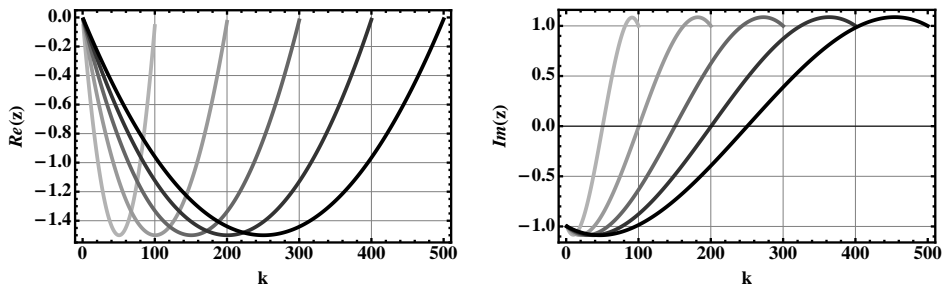}
\caption{Real and imaginary parts of the zeros of the normalized Bessel polynomials $y_n(x/n)$ for $n=100, 200, 300, 400$, and $500$, plotted in gray-level intensity, from lower to higher, according to the value of $n$.}
\end{figure}
Thus, we propose the following expressions
\begin{eqnarray}\label{zerpol}
\tilde{x}(k,n)&=&a_2(n) k^2+a_1(n) k+a_0(n),\\
\tilde{y}(k,n)&=&b_3(n) k^3+b_2(n) k^2+b_1(n) k+b_0(n)\nonumber
\end{eqnarray}
to fit our data. To find the relationship between the coefficients of these polynomials and $n$, we take into account the numerical behavior of the data at the middle
and end points. According to this, $\tilde{x}(k,n)$ and $\tilde{y}(k,n)$ can be determined by
\begin{equation}\label{conre}
\tilde{x}(0,n)=0,\quad \tilde{x}(\frac{n}{2},n)=-\frac{3}{2 n},\quad \tilde{x}(n+1)=0,
\end{equation}
and
\begin{equation}\label{conim}
\tilde{y}(1,n)=-\frac{1}{n},\quad \tilde{y}(\frac{n+1}{2},n)=0,\quad \frac{d\tilde{y}(k,n)}{dk}\Big\vert_{k=\frac{n}{2}}=\frac{4}{n^2},\quad \tilde{y}(n,n)=\frac{1}{n}.
\end{equation}
These conditions lead to the following coefficients
\begin{eqnarray}\label{coepol}
a_2(n)&=&\frac{6}{n^2 (n+2)},\qquad a_1(n)=-\frac{6 (n+1)}{n^2 (n+2)},\qquad a_0(n)=0,\nonumber\\
b_3(n)&=&-\frac{8 (n-2)}{n^2 \left(n^3-3 n^2+2\right)},\quad b_2(n)=\frac{12 (n-2) (n+1)}{(n-1) n^2 \left(n^2-2 n-2\right)},\\
b_1(n)&=&-\frac{2 \left(n^3+6 n^2-12 n-4\right)}{(n-1) n^2 \left(n^2-2 n-2\right)},\quad b_0(n)=-\frac{n^3-5 n^2+6}{n \left(n^3-3 n^2+2\right)}.\nonumber
\end{eqnarray}
The substitution of (\ref{coepol}) in (\ref{zerpol}) yields approximate closed-form expressions 
\begin{equation}\label{zktil}
\tilde{z}_k=\tilde{x}(k,n)+ i \tilde{y}(k,n),
\end{equation}
$k=1,2,\cdots,n$, that converge to the zeros $z_k$ of the Bessel polynomial $y_n(x)$, as we will show in the following.
\section{Convergence}\label{conv}
Following \cite{Olv54}, we define 
\begin{equation}\label{weq}
W(z)=\frac{e^{\sqrt{1+1/z^2}}}{z(1+\sqrt{1+1/z^2})}
\end{equation}
and denote by $\Gamma$ the curve defined by 
\[
\Gamma=\{z\in{\mathbb C}:\vert W(z)\vert=1\quad\text{and}\quad \vert\text{arg} z\vert\ge\frac{\pi}{2}\},
\]
which contains the limit points $\hat{\omega}_k$ of the zeros of the normalized Bessel polynomial $y_n(x/n)$. Then, it has been proved in \cite{Car92} that the zero $\omega_k$ of $y_n(x/n)$ approaches to order $O(1/n)$ the limit value $\hat{\omega}_k$, i.e,
\begin{equation}\label{difzkzgo}
\vert \omega_k-\hat{\omega}_k\vert=O(1/n),
\end{equation}
as $n\to\infty$.\\
Thus, if we show that $\vert  \tilde{\omega}_k-\hat{\omega}_k\vert=O(1/n)$, we will have proved that 
\begin{equation}\label{difzkzt}
\vert \omega_k-\tilde{\omega}_k\vert=O(1/n),
\end{equation}
and therefore, taking into account that $\omega_k=n z_k$, the explicit expression (\ref{zktil}) approaches to order $O(1/n^2)$ the zero $z_k$ of the Bessel polynomial $y_n(x)$.\\
To this purpose, we simply substitute $\tilde{\omega}_k=n \tilde{z}_k$ in (\ref{weq}) to obtain, after a lengthily calculation, that the expansion of  $W(\tilde{\omega}_k)$ in terms of $1/n$ is
\begin{eqnarray*}
W(\tilde{\omega}_k)&=&1-2 \bigg[\sqrt{10 k^2-2 k+1} \sin ^2\left(\frac{6 k}{4 k-4}\right)\\
&-&\sqrt{10 k^2-2 k+1} \cos ^2\left(\frac{6 k}{4k-4}\right)+k-1\bigg]\frac{1}{n}+O\left(\frac{1}{n^{3/2}}\right),
\end{eqnarray*}
and this implies that
\[
\vert W(\tilde{\omega}_k)\vert=1+O\left(\frac{1}{n}\right)
\]
for fixed $k$. Thus, $\tilde{\omega}_k$ approaches to order $O(1/n)$ the $\Gamma$ curve and (\ref{difzkzt}) follows. From here we have that
\begin{equation}\label{oxkxtk}
\vert z_k-\tilde{z}_k \vert=O(1/n^2)
\end{equation}
as $n\to\infty$. Numerical calculations confirm and extend this result. Figure 2 shows the behavior of the maxima of $\vert z_k-\tilde{z}_k\vert$ over $k$ as they depend on $n$.  The numbers computed by (\ref{elein}) are taken as the exact zeros $z_k$. The displayed data shows numerical uniform convergence on the values of $k$, not only for fixed $k$. A fit of these data gives $1/n^a$ with $a=1.7$.
\begin{figure}[H]\label{Figdos}
\centering
\includegraphics[scale=0.8]{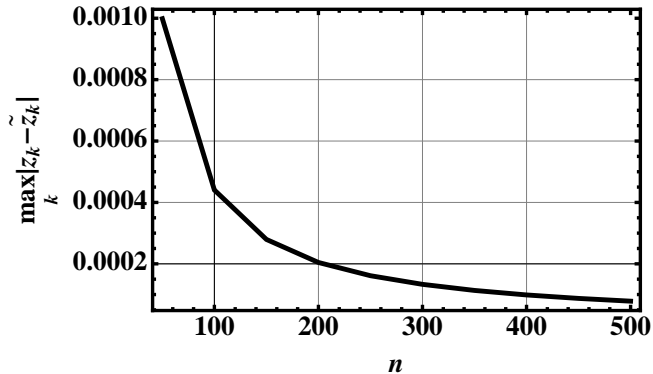}
\caption{Plot of the values of $\max_{k=1}^n\vert z_k-\tilde{z}_k\vert$ against $n$.}
\end{figure}
\noindent
\section{Some few tests}\label{misc}
Just to give examples of the application of the approximate expression (\ref{zktil}), we consider the following cases.
\subsection{The real zero}
An closed-form formula for the unique real zero $\alpha(n)$ of the Bessel polynomial $y_n(x)$ can be obtained by the substitution of $k=(n + 1)/2$ in the real part of (\ref{zktil}), $\tilde{x}(k,n)$. This gives
\begin{equation}\label{alfcam}
\alpha(n)=-\frac{3 (n+1)^2}{2 n^2 (n+2)} +O(1/n^2).
\end{equation}
as our new result. In \cite{Gro78, Bru81} asymptotic expressions for $\alpha(n)$ are given. Particularly, the following formula
\begin{equation}\label{alfvar}
\alpha(n)=-\frac{2}{1.32549 n + 0.662743}+O(1/n),
\end{equation}
can be obtained from a more general expression given in \cite{Bru81}. Deleting the $O$-terms and expanding both results in powers of $1/n$ we find that
\[
\tilde{\alpha}(n)\simeq-\frac{3}{2n},\qquad \alpha(n)\simeq-1.50888 \frac{1}{n},
\]
indicating good agreement between the two approaches.
\subsection{Power sums} Here we carry out the corresponding multiplications and use some cases of Faulhaber's formula. Then we compare our results with the exact ones.
\\
1. {\sl Sum of the zeros}. The simple sum of $\tilde{z}_k$ [cf. (\ref{zktil})], gives
\[
\tilde{s}_1(n)=\sum_{k=1}^n\tilde{z}_k=-\frac{n+1}{n}=-1+O(1/n).
\]
The exact result is $s_1(n)=-1$, as can be seen from (\ref{bespoy}). 
\\
2.  {\sl Sum of the squares of the zeros}. In this case we obtain
\[
\tilde{s}_2(n)=\sum_{k=1}^n\tilde{z}_k^2=\frac{p_2(n)}{q_2(n)}
\]
where
\begin{eqnarray*}
p_2(n)&=&55 n^8+15 n^7-800 n^6-612 n^5+4064 n^4+1740 n^3\\
&-&1696 n^2-2832 n-3312,\\
q_2(n)&=&105 n^3 \left(n^2-2 n-2\right)^2\left(n^2+n-2\right).
\end{eqnarray*}
The exact sum 
\[
s_2(n)=\frac{1}{2 n - 1},
\]
can be found elsewhere \cite{Gal84}. Expanding both expressions in powers of $1/n$ we find that
\[
\tilde{s}_2(n)=\frac{11}{21n}+O(1/n^2),\qquad s_2(n)=\frac{1}{2 n}+O(1/n^2).
\]
3. {\sl Sum of the cubes of the  zeros}. In a similar form, we obtain that
\[
\tilde{s}_3(n)=\sum_{k=1}^n\tilde{z}_k^3=\frac{p_3(n)}{q_3(n)}
\]
where
\begin{eqnarray*}
p_3(n)&=&-n^{10}+28 n^9-217 n^8+468 n^7+1002 n^6-3804 n^5-1076 n^4+5936 n^3\\
&-&3848 n^2+12816 n+19584,\\
q_3(n)&=&35 (n-1) n^5 \left(n^3-6 n-4\right)^2.
\end{eqnarray*}
In this case, the exact sum is zero  \cite{Gal84}. Expanding $\tilde{s}_3(n)$ in $1/n$ we find that
\[
\tilde{s}_3(n)=O(1/n^2).
\]
\section{Final comment}
Note that the approximate formula for $z_k$ given above is not unique. There exist other functions to fit the zeros obtained through the electrostatic equations (\ref{elein}), and there are other conditions to impose at the extreme and middle points of the fitting interval. For instance, the imaginary part $\tilde{y}(k,n)$, can be fitted by a polynomial of degree 5, but this does not improve the rate of convergence and, on the other hand, the calculations become more lengthy. 
\section{Acknowledgment}
The authors thank Consejo Nacional de Ciencia y Tecnolog\'{\i}a for the financial support given to this project.

\vskip1cm


\begin{thebibliography}{99}
\bibitem{Kra49} H.L. Krall and O. Frink, {\sl A New Class of Orthogonal Polynomials: The Bessel Polynomials}, Trans. Amer. Math. Soc., {\bf 65} (1949) 100-115.
\bibitem{Gro78}  E. Grosswald, {\sl Bessel Polynomials}, Lecture Notes in Mathematics, Vol 698, Springer-Verlag, Berlin, 1978.
\bibitem{Sri84} H. M. Srivastava, {\sl Some Orthogonal Polynomials Representing the Energy Spectral Functions for a Family of Isotropic Turbulence Fields}, Zeitschr. Angew. Math. Mech. {\bf 64} (1984) 255-257.
\bibitem{Lop11} J.L. L\'opez and N.M. Temme, {\sl Large degree asymptotics of generalized Bessel polynomials}, J. Math. Anal. Appl. {\bf 377} (2011) 30-42.
\bibitem{Ege10} \"O. E\u{g}ecio\u{g}lu, {\sl Bessel Polynomials and the Partial Sums of the Exponential Series} SIAM J. Discrete Math. {\bf 24} (2010) 1753Ð1762.
\bibitem{Ber08} C. Berg and C. Vignat {\sl Linearization coefficients of Bessel polynomials and properties of Student-t distributions}, Constr. Approx., {\bf 27} (2008) 15-32.
\bibitem{Pas00} L. Pasquini, {\sl Accurate computation of the zeros of the generalized Bessel polynomials}, Numer. Math. {\bf 86} (2000) 507Ð538.
\bibitem{Car92} A.J. Carpenter, {\sl  Asymptotics for the zeros of the generalized Bessel polynomials}, Numer. Math. {\bf 62} (1992) 465-482.
\bibitem{Olv54} F.W.J. Olver, {\sl The asymptotic expansions of Bessel functions of large order}, Phil. Trans. R. Soc. Lon. A, {\bf 247} (1954) 338-368.
\bibitem{Run83} H.J. Runckel, {\sl Zero-free parabolic regions for polynomials with complex coefficients}, Proc. Amer. Math. Soc, {\bf 88} (1983) 299-304.
\bibitem{Bru81} M.G. de Bruin, E.B. Saff, R.S. Varga, {\sl On the zeros of generalized Bessel polynomials I, II}, Indag. Math. {\bf 84} (1981) 1-25
\bibitem{Gal84} F. G\'alvez and J.S. Dehesa, {\sl Some open problems of generalised Bessel polynomials}, J. Phys. A, {\bf 17} (1984) 2759-2766.
\bibitem{Pit99} J. Pitman, {\sl A lattice path model for the Bessel polynomials}, Technical report 551, Dept. Statistics, University of California, Berkeley, USA, 1999, {\it http://www.stat.berkeley.edu/users/pitman/}
\bibitem{Bia07} G. Bianchi and R. Sorrentino, {\sl Electronic filter simulation and design}, McGrawÐHill, New York, NY, 2007.
\bibitem{Sze75} G. Szeg\H{o}, {\sl Orthogonal Polynomials}, Colloquium Publications, American Mathematical Society, Providence, Rhode Island, 1975.
\bibitem{Mar05} F. Marcell\'an, A. Mart\'{\i}nez-Finkelshteinb, and P. Mart\'{\i}nez-Gonz\'alez, {\sl Electrostatic models for zeros of polynomials: old, new, and some open problems},  J. Comp. Appl. Math. {\bf 207} (2007) 258-272.
\bibitem{Cam99} R.G. Campos, {\sl Perturbed zeros of orthogonal polynomials}, Bol. Soc. Mat. Mexicana. 5 (1999) 143-153.
\bibitem{Cam97} R.G. Campos, {\sl Solving nonlinear two point boundary value problems}, Bol. Soc. Mat. Mexicana. 3 (1997) 279-297
\bibitem{Cam95} R.G. Campos y L.A. Avila, {\sl Some properties of orthogonal polynomials satisfying fourth order differential equations}, Glasgow Math. J., {\bf 37} (1995) 105-113


\end{thebibliography}
\end{document}